\documentclass[useAMS,usenatbib]{mn2e}
\usepackage{lscape,graphicx,amssymb,aas_macros,bm}
\usepackage{amsmath}
\usepackage{indentfirst}
\usepackage{pdflscape}
\usepackage{placeins}
\usepackage{nicefrac}
\usepackage{afterpage}
\usepackage{booktabs}
\usepackage{color}
\usepackage{float}
\usepackage{rotating}
\usepackage{adjustbox}

\newcommand{\ewha}{\mbox{${\rm EW}({\rm H}\alpha)$}}
\newcommand{\apg}{\:^{>}_{\sim}\:}
\newcommand{\apl}{\:^{<}_{\sim}\:}

\newcommand{\etal}{et al.}
\newcommand{\HI}{{\mbox{H\,{\scriptsize I}}}}

\newcommand{\kms}{\mbox{km\ s${^{-1}}$}}
\newcommand{\lya}{\mbox{${\rm Ly}\alpha$}}

\newcommand{\CIII}{{\mbox{C\,{\scriptsize III}}}}

\newcommand{\SiIII}{{\mbox{Si\,{\scriptsize III}}}}
\newcommand{\OVI}{{\mbox{O\,{\scriptsize VI}}}}

\newcommand{\OIII}{{\mbox{[O\,{\scriptsize III}]}}}

\newcommand{\mstar}{{\mbox{$M_{\rm star}$}}}
\newcommand{\msun}{{\mbox{M$_{\odot}$}}}
\newcommand{\sfr}{{\mbox{M$_{\odot}$\,yr$^{-1}$}}}

\title[CGM in Massive Quiescent Halos III]{Characterizing circumgalactic gas around massive ellipticals at $\bm{z}\approx 0.4$ III.\ The galactic environment of a chemically-pristine Lyman limit absorber}

\author[Chen et al.]{Hsiao-Wen Chen$^{1}$\thanks{E-mail: hchen@oddjob.uchicago.edu},
Sean D.\ Johnson$^{2,3}$\thanks{Hubble \& Carnegie-Princeton Fellow}, Lorrie A.\ Straka$^{4}$, Fakhri S.\ Zahedy$^{1}$, \newauthor Joop Schaye$^{4}$, Sowgat Muzahid$^{4}$, Nicolas Bouch\'e$^{5,6}$, Sebastiano Cantalupo$^{7}$, \newauthor Raffaella Anna Marino$^{7}$, Martin Wendt$^{8,9}$
\\ 
$^{1}$Department of Astronomy \& Astrophysics, The University of Chicago, Chicago, IL 60637, USA \\
$^{2}$Department of Astrophysics, Princeton University, Princeton, NJ 08544, USA \\
$^{3}$The Observatories of the Carnegie Institution for Science, 813 Santa Barbara Street, Pasadena, CA 91101, USA \\
$^{4}$Leiden Observatory, Leiden University, PO Box 9513, NL-2300 RA Leiden, the Netherlands\\
$^{5}$CNRS/IRAP, 9 Avenue Colonel Roche, F-31400 Toulouse, France\\
$^{6}$Univ Lyon, Univ Lyon1, Ens de Lyon, CNRS, Centre de Recherche Astrophysique de Lyon UMR5574, F-69230, Saint-Genis-Laval, France \\
$^{7}$Department of Physics, ETH Wolfgang$-$Pauli$-$Strasse 27, 8093, CH-8093 Z\"urich, Switzerland \\
$^{8}$Institut f\"ur Physik und Astronomie, Universit\"at Potsdam,Karl-Liebknecht-Str. 24$/$25, 14476 Golm, Germany\\
$^{9}$Leibniz-Institut f\"ur Astrophysik Potsdam (AIP), An der Sternwarte 16, 14482 Potsdam, Germany\\
}

\begin{document}

\pagerange{\pageref{firstpage}--\pageref{lastpage}} \pubyear{2017}

\maketitle

\label{firstpage}

\begin{abstract}

This paper presents a study of the galactic environment of a
chemically-pristine ($<0.6$\% solar metallicity) Lyman Limit system
(LLS) discovered along the sightline toward QSO
SDSSJ\,135726.27$+$043541.4 ($z_{\rm QSO}=1.233$) at projected
distance $d=126$ physical kpc (pkpc) from a luminous red galaxy (LRG)
at $z=0.33$.  Combining deep {\it Hubble Space Telescope} images, MUSE
integral field spectroscopic data, and wide-field redshift survey data
has enabled an unprecedented, ultra-deep view of the environment
around this LRG-LLS pair.  A total of 12 galaxies, including the LRG,
are found at $d\,\apl\,400$ pkpc and line-of-sight velocity
$\Delta\,v<600$ \kms\ of the LLS, with intrinsic luminosity ranging
from $0.001\,L_*$ to $2\,L_*$ and a corresponding stellar mass range
of $\mstar\approx 10^{7-11}\,\msun$.  All 12 galaxies contribute to a
total mass of $\mstar=1.6\times 10^{11}\,\msun$ with $\approx 80$\%
contained in the LRG.  The line-of-sight velocity dispersion of these
galaxies is found to be $\sigma_{\rm group}=230$ \kms\ with the center
of mass at $d_{\rm group}=118$ pkpc and line-of-sight velocity offset
of $\Delta\,v_{\rm group}=181$ \kms\ from the LLS.  Three of these
are located at $d\apl 100$ pkpc from the LLS, and they are all faint
with intrinsic luminosity $<0.02\,L_*$ and gas phase metallicity of
$\approx 10$\% solar in their interstellar medium.  The disparity in
the chemical enrichment level between the LLS and the group members
suggests that the LLS originates in infalling intergalactic medium and
that parts of the intergalactic gas near old and massive galaxies can
still remain chemically pristine through the not too distant past.

\end{abstract}


\section{Introduction}

Luminous red galaxies (LRGs) are selected based on their photometric
and spectroscopic properties that resemble nearby elliptical galaxies
(e.g., Eisenstein \etal\ 2001).  These galaxies comprise primarily old
stars with little/no on-going star formation (e.g., Roseboom
\etal\ 2006; Gauthier \& Chen 2011) and represent more than 90\% of
massive galaxies with stellar mass $\mstar\apg 10^{11}\,\msun$ (e.g.,
Peng \etal\ 2010; Tinker \etal\ 2013) with a corresponding dark matter
halo mass of $M_h>10^{13}\,\msun$ (Blake \etal\ 2008).  Therefore,
they offer an ideal laboratory for studying feeding and feedback in
massive halos that are generally free of complex starburst driven
winds (see Chen 2017 for a recent review).

Motivated by previous studies that reported abundant
chemically-enriched cool gas in LRG halos (e.g., Gauthier \etal\ 2009,
2010, Lundgren \etal\ 2009; Gauthier \& Chen 2011; Bowen \& Chelouche
2011; Zhu \etal\ 2014; Huang \etal\ 2016), we are carrying out a UV
absorption-line survey to study the circumgalactic medium (CGM) around
16 LRGs using the Cosmic Origins Spectrograph (COS) on board the {\it
  Hubble Space Telescope} ({\it HST}) (Chen \etal\ 2018; hereafter
Paper I).  The LRGs in this COS-LRG sample were selected uniformly
with stellar mass $\mstar>10^{11}\,\msun$ and no prior knowledge of
the presence/absence of any absorption features.  The program was
designed to enable accurate measurements of the neutral hydrogen
column density $N(\HI)$ based on observations of the full Lyman
series.  This COS-LRG sample therefore allows a systematic study of
the ionization state and chemical enrichment of the CGM in massive
quiescent halos.

The main findings from our initial analysis presented in Paper I can
be summarized as follows.  First, high-$N(\HI)$ gas is common in these
massive quiescent halos with a median of $\langle\,\log\,N(\HI)/{\rm
  cm}^{-2}\rangle = 16.6$ at projected distances $d\apl 160$ physical
kpc (pkpc).  For all but three LRGs the associated Lyman absorption
series is detected.  The mean covering fraction of optically-thick
absorbers with $\log\,N(\HI)/{\rm cm}^{-2}\apg 17.2$ is found to be
$\langle\kappa\rangle_{\rm 17.2}=0.44^{+0.12}_{-0.11}$ at $d\le 160$
pkpc and $\langle\kappa\rangle_{\rm 17.2}=0.71^{+0.11}_{-0.20}$ at
$d\apl 100$ pkpc.  Secondly, the gas has been enriched with heavy
elements with \CIII\,$\lambda\,977$ and \SiIII\,$\lambda\,1206$
absorption lines being the most dominant features.  The observed high
covering fraction of \CIII\ absorbing gas in LRG halos is
statistically consistent with what is seen in COS-Halos blue galaxies.
The ``bimodality'' observed in \OVI-absorbing gas between star-forming
and quiescent galaxies (e.g., Chen \& Mulchaey 2009; Tumlinson
\etal\ 2011) is not seen in these intermediate ions.  Finally, while
the line-of-sight velocity separations between the \HI\ absorbing gas
and LRGs are characterized by a mean and dispersion of
$\langle\,v_{{\rm gas}-{\rm LRG}}\rangle=29$ \kms\ and
$\sigma_{(v_{{\rm gas}-{\rm LRG}})}=171$ \kms, eight of the 13 LRG
halos with detected absorption features were resolved into multiple
components with velocities separated by as much as $\approx 360$ \kms.

In a follow-up study, Zahedy \etal\ (2018; hereafter Paper II)
performed a detailed component by component analysis that
simultaneously incorporated multiple ionic transitions for
constraining the density, $n_{\rm H}$, and metallicity, [M/H], of the
CGM in LRG halos based on comparisons with a grid of Cloudy (Ferland
\etal\ 2013; v13.03) ionization models.
An interesting finding of Paper II
is that the derived gas metallicities of individual components can
vary by more than an order of magnitude in a single LRG halo, from
$<1/10$ solar to solar or super solar irrespective of the adopted
ionizing spectrum.  This applies to gas both at $d<100$ pkpc and
beyond.  In addition, solar-metallicity gas is observed out to $d=160$
pkpc in half of the COS-LRG sample, while metal-poor gas of
metallicity $\apl 1/100$ solar is also found in at least three cases.
Similarly, the inferred gas densities vary by two orders of magnitude
between individual components in a single halo, from $n_{\rm H}\approx
0.1\,{\rm cm}^{-3}$ to $n_{\rm H}\apl 0.001\,{\rm cm}^{-3}$, while the
mean gas density averaged over all components is $\langle\,n_{\rm
  H}\,\rangle\approx 0.005-0.01\,{\rm cm}^{-3}$ for the majority of
LRG halos at $d=40-160$ pkpc.

In one particular LRG halo, SDSSJ\,135727.27$+$043603.3 at $z=0.3295$,
we detected a strong absorber of $\log\,N(\HI)/{\rm cm}^{-2}=17.48\pm
0.01$\footnote{The error quoted above represents only statistical
  uncertainties.  Systematic uncertainties due to continuum placement
  errors are not included.  For comparison, Lehner et al.\ (2018)
  reports $\log\,N(\HI)/{\rm cm}^{-2}=17.55\pm 0.05$ for this
  system.}, a Lyman limit system (LLS), at $d=126$ pkpc and $z_{\rm
  LLS}=0.3287$ which yields a line-of-sight velocity offset of
$v_{{\rm gas}-{\rm LRG}}\approx -180$ \kms.  This LLS is characterized
by a single component of Doppler parameter $b_\HI=18$ \kms\ and no
associated metal ion features down to sensitive column density limits
(Figure 1).  The observed simple absorption profiles of the hydrogen
lines place a strong constraint on the gas temperature at $T< 2\times
10^4$ K.  A 95\% upper limit to the gas metallicity of
[M/H]\,$<-2.2$ was obtained for the LLS
based on the non-detections of any metal transitions and a size limit
of $\apl 1$ kpc for the absorber (Paper II).  The low metallicity
limit is also confirmed by Lehner et al.\ (2018), who report $[{\rm
    Mg}/{\rm H}]=-2.3\pm0.2$ for this absorber.  Not only is the
  observed metal content significantly lower than the typical solar
  metallicity found in nearby elliptical galaxies (e.g., Mathews \&
  Brighenti 2003) or the near-solar metallicity of the interstellar
  medium (ISM) of an intermediate-redshift elliptical (Zahedy
  \etal\ 2017), it is also among the most metal-poor LLSs found at $z
  < 1$ (e.g., Ribaudo et al.\ 2011; Lehner \etal\ 2018).  The observed
  chemically-pristine gas near a massive quiescent galaxy strongly
  suggests that the LLS originates in infalling gas clouds from the
  intergalactic medium (IGM).

\begin{figure}
 \includegraphics[scale=0.45]{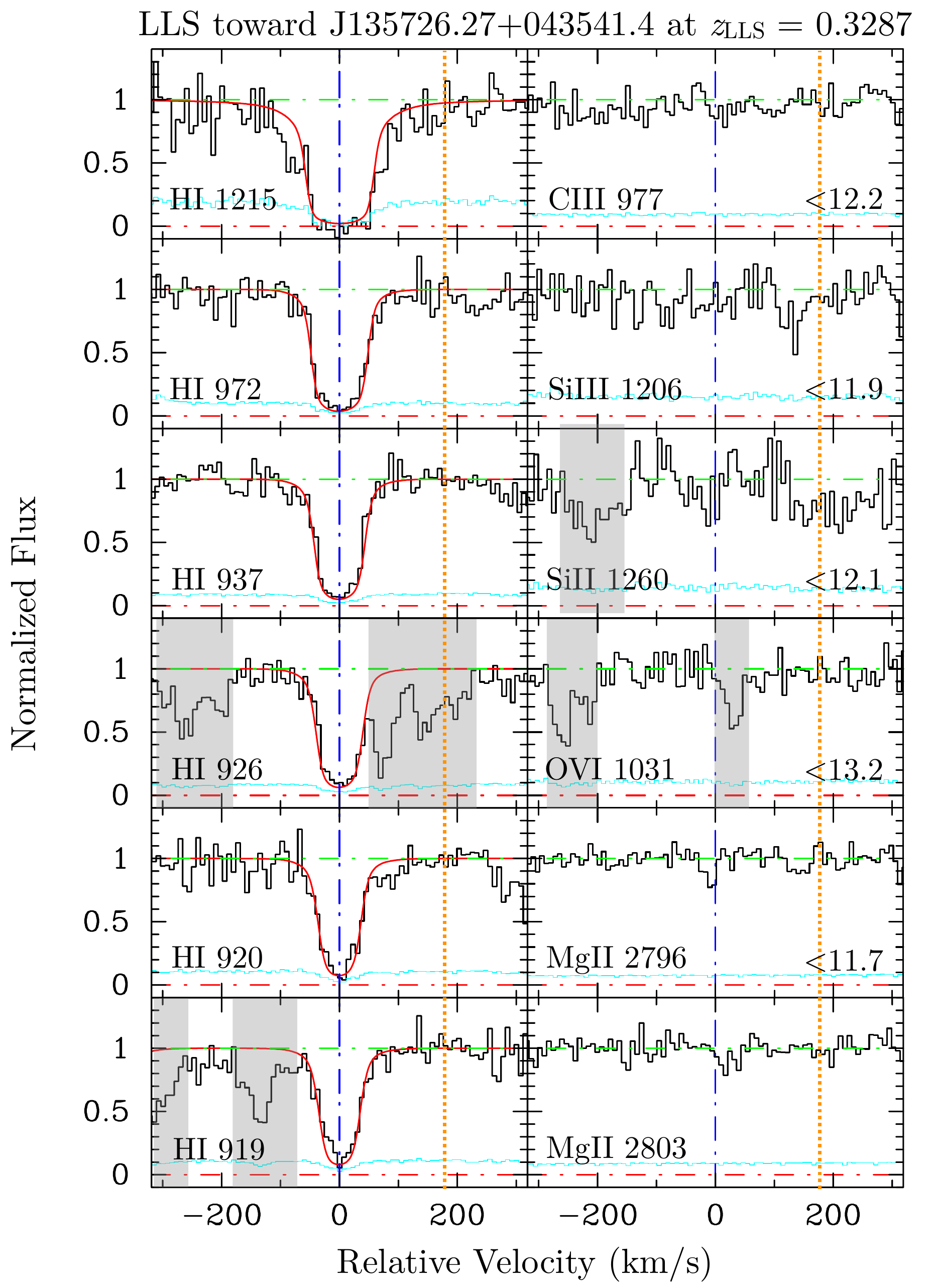}
 \caption{A subset of absorption transitions of the LLS with
   $\log\,N(\HI)/{\rm cm}^{-2}=17.48\pm 0.01$ at $d=126$ pkpc from LRG
   SDSSJ\,135727.27$+$043603.3 at $z_{\rm LRG}=0.3295$ (see Chen
   \etal\ 2018 and Zahedy \etal\ 2018 for a complete list).
   Contaminating features are greyed out for clarity.  Zero relative
   velocity corresponds to the redshift of the LLS at $z_{\rm
     LLS}=0.3287$.  The vertical dotted line marks the relative
   line-of-sight velocity of the LRG at $+180$ \kms.  The hydrogen
   Lyman series lines displayed in the left panels are characterized
   by a single component of $b_\HI=18$ \kms, constraining the gas
   temperature to be $T< 2\times 10^4$ K.  In contrast, no heavy ions
   are detected to sensitive upper limits (right panels).  The 95\%
   upper limits to the underlying ionic column densities in
   logarithmic values are shown in the lower-right corner of the right
   panels.} \label{figure:absspec}
\end{figure}

Here we present deep wide-field integral field spectroscopic data from
the MUSE-QuBES program (PI: Schaye; see also Segers \etal\ 2018 \&
Straka \etal\ 2018, in preparation), wide-field multi-object
spectroscopic data obtained using the Low Dispersion Survey
Spectrograph 3 (LDSS3) on the Magellan Clay Telescope, and exquisite
{\it Hubble Space Telescope} ({\it HST}) imaging data (PI: Straka) of
the field.  These imaging and spectroscopic data together provide a
high-definition view of the galaxy environment around the
chemically-pristine LLS and the LRG at $z=0.33$, and offer valuable
insights into the origin of chemically-pristine gas observed near a
massive, quiescent galaxy.  The paper is organized as follows.  In
Section 2, we describe the imaging and spectroscopic observations, and
the construction of photometric and spectroscopic catalogs of faint
galaxies in the field.  In Section 3, we present properties of all
galaxies identified in the vicinity of the chemically-pristine LLS at
$z=0.33$.  In Section 4, we discuss the implications of our findings.
Throughout the paper, we adopt $12 + \log({\rm O}/{\rm H}) = 8.69\pm
0.05$ for the solar oxygen abundance (Asplund \etal\ 2009), and a
standard $\Lambda$ cosmology, $\Omega_M$ = 0.3 and
$\Omega_\Lambda$=0.7 with a Hubble constant $H_{\rm 0} = 70\rm
\,km\,s^{-1}\,Mpc^{-1}$.

\section{Observations and Data Analysis}

To carry out a deep galaxy survey around the LRG and the associated
chemically-pristine LLS at $z=0.33$, we have assembled deep imaging
and spectroscopic data both from our own observations and from the
public archives.  Here we describe the observations and data analysis.

\begin{figure*}
\includegraphics[width=\textwidth]{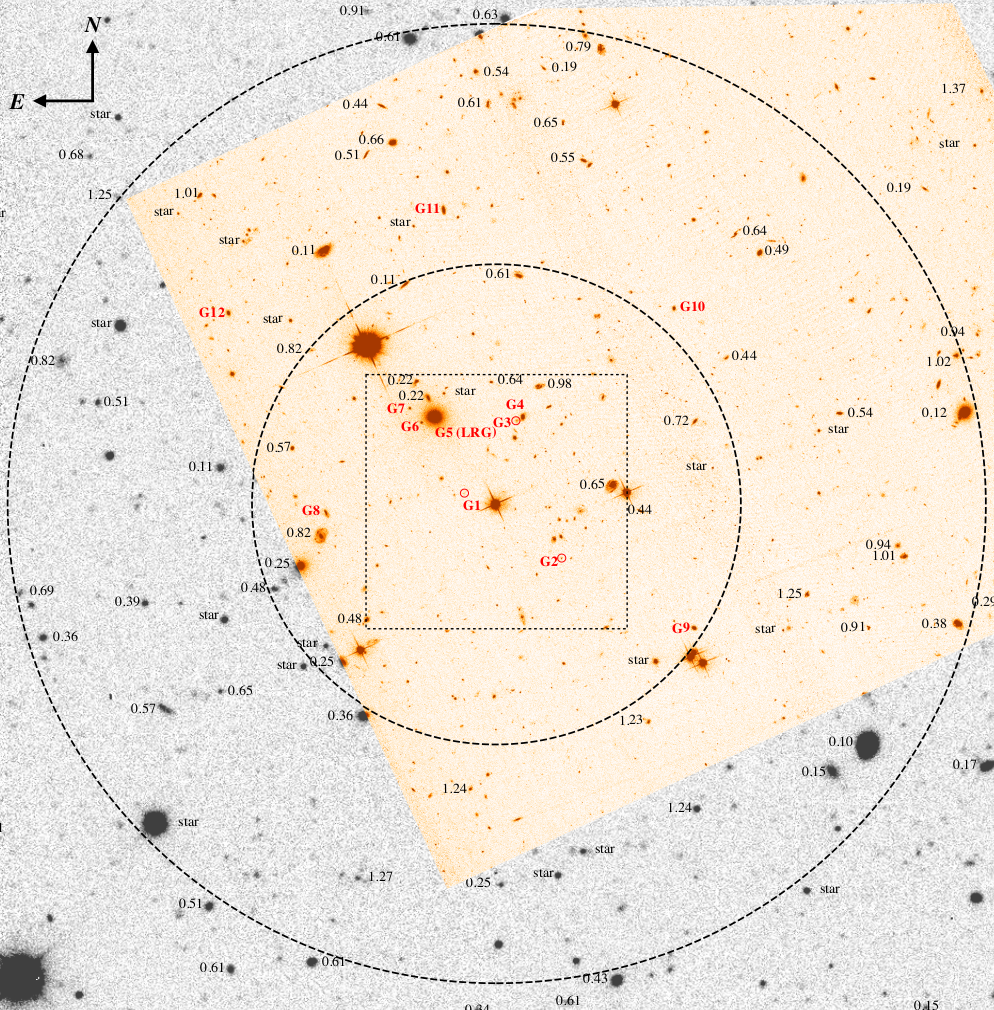}
\caption{The field centered at QSO SDSSJ\,135726.27$+$043541.4 at
  $z_{\rm QSO}=1.233$ observed in DECaLS $r$-band (grey-scale image)
  and in {\it HST}/ACS and the F814W filter (displayed in heat map).
  The dashed circles mark the $1'$ and $2'$ radii from the QSO
  sightline.  The dotted square box marks the field of view of MUSE
  observations (see \S\ 2.2 and Figure 3 below).  Spectroscopically
  identified objects from the LDSS3 survey are marked with redshift
  measurements next to them (see \S\ 2.3).  Galaxies associated with
  the pristine LLS at $z=0.33$ from the combined LDSS3 and MUSE survey
  are labeled G1 through G12 in red.  The DECaLS image reaches a
  5-$\sigma$ limiting magnitudes of $AB(r)=24.4$ for extended sources.
  The HST/ACS image (PID$=$14660; PI: Straka) reaches a 5-$\sigma$
  limiting magnitude of $AB({\rm F814W})=27.2$ over a $0.5''$-diameter
  aperture, providing resolved morphologies for DECaLS identified
  objects and revealing fainter objects missed in the DECaLS imaging
  data.  The previously identified LRG is G5 at $\approx 27''$
  ($d=126$ pkpc) northeast of the QSO at $z=0.3295$, surrounded by two
  other compact and fainter neighbors at similar redshifts (see Tables
  1 \& 2 below, and Figure 3 for a zoom-in view).  Combining deep
  imaging and spectroscopic data has uncovered an overdensity of faint
  galaxies at $d<400$ pkpc from the pristine LLS at $z=0.33$.}
\end{figure*}

\subsection{Imaging observations and the object catalog}

The field around QSO, SDSSJ\,135726.27$+$043541.4, at $z_{\rm
  QSO}=1.233$ is covered by the public DECam Legacy Survey (DECaLS;
PI: Schlegel and Dey; http://legacysurvey.org/decamls/; see also Dey
\etal\ 2018), which images equatorial fields in the $g$-, $r$-, and
$z$-bands.  Between five and eight individual exposures in each of the
$g$-, $r$-, and $z$-bands were obtained for this field, reaching
5-$\sigma$ imaging depths of $AB(g)=25.1$, $AB(r)=24.4$ and
$AB(z)=23.6$ mag for extended sources.  Coadded $g$-, $r$-, and
$z$-band images of the field and the associated object catalog were
retrieved from the DECaLS data release 5 website.  The mean seeing
conditions in the coadded images are characterized by a
full-width-at-half-maximum (FWHM) for point sources of $1''$, $0.9''$,
and $0.8''$ in the $g$-, $r$-, and $z$-band, respectively.  The
available $g$-, $r$-, and $z$-band photometric measurements provide
constraints for the broad-band spectral energy distributions of
relatively bright objects.  The stacked $r$-band image is presented in
Figure 2.

We have also obtained high-quality {\it Hubble Space Telescope} ({\it
  HST}) images of the field using the Wide Field Channel (WFC) in the
Advanced Camera for Surveys (ACS) and the F814W filter (PID$=$14660;
PI: Straka).  The observations were carried out on 2017 July 5
UT. Four exposures of 543 s each were obtained and processed using the
standard {\it HST} reduction pipeline.  Individual frames were
corrected for geometric distortion using drizzle, registered to a
common origin, filtered for deviant pixels, and combined to form a
final stacked image.  The final stacked F814W image covers a region of
roughly $3.5'\times 3.5'$ and reaches a 5-$\sigma$ limiting magnitude
of $AB({\rm F814W})=27.2$ over a $0.5''$-diameter aperture,
significantly deeper than the public $g$-, $r$-, and $z$-band images
from DECaLS.  The F814W image replaces the DECaLS $r$-band image in
Figure 2 in the overlapping region.  In addition to resolving galaxy
morphologies, the {\it HST} ACS/F814W image uncovers more than twice
as many faint galaxies as found in the DECaLS images.

While the greater depth and higher spatial resolution of the {\it
  HST}/ACS F814W image make the F814W an ideal choice for detections
of faint galaxies, the available DECaLS data provide a wider field
coverage for studying the large-scale environment.  To construct the
object catalog, we first adopted the public photometric catalog from
DECaLS.  For the area that is also covered by the {\it HST}/ACS F814W
image, we performed source detections in the F814W image using the
SExtractor program (Bertin \&Arnouts 1996) and identified additional
faint objects with $AB({\rm F814W})<27$ mag.  We then combined the
DECaLS and F814W catalogs to form a master catalog by matching object
coordinates.  Nearly all objects with $AB({\rm F814W})<23.5$ mag have
an entry in the DECaLS catalog.  For objects that are not detected in
the DECaLS images, the 5-$\sigma$ limiting magnitudes computed for
extended sources in this region by the DECaLS team are adopted (see
e.g., Dey \etal\ 2018).

\subsection{MUSE integral field spectroscopic observations}

Wide-field integral field spectroscopic data of the QSO field have
been obtained using the Multi-Unit Spectroscopic Explorer (MUSE; Bacon
et al.\ 2010), as part of the MUSE Quasar-field Blind Emitter Survey
(MUSE-QuBES; PI: Schaye).  MUSE observes a field of $1'\times 1'$ at a
pixel scale of $0.2''$ and a spectral resolution of ${\rm FWHM}\approx
120$ \kms\ at 7000 \AA.  It enables a deep search of faint galaxies
within 140 pkpc in projected distance of the QSO sightline at
$z=0.33$.  The observations were carried out in May of 2015, in blocks
of one hour each for a total of two hours.  Each observing block was
split into four exposures of 900 s each.  Small dithers of a few
arcseconds in the telescope pointing and offsets in the position angle
in steps of 90 degrees were applied between exposures to smooth out
the tiling pattern in the combined data cube.  Individual exposures
were first reduced using the standard MUSE data reduction pipeline
(v1.2; Weibacher \etal\ 2012) and further processed using the
CubExtractor package (Cantalupo et al., in prep; see also Borisova
\etal\ 2016; Marino \etal\ 2018), which include bias subtraction, flat-fielding,
illumination correction, and wavelength and flux calibrations. (see
Segers \etal\ 2018 for a detailed description).

Within the MUSE $1'\times 1'$ field of view, we identified 110 objects
of $AB({\rm F814W})<27$ mag, of which 47 are identified in the DECaLS
photometric catalog.  All but one of the remaining 63 objects are
faint with $AB({\rm F814W})>23.5$ mag and do not have an entry in the
DECaLS catalog.  The only exception is J135727.50$+$043601.98 (G6 in
Figures 2 \& 3) with $AB({\rm F814W})=22.35$ mag, which occurs at
$\approx 3''$ from the LRG (G5) with $AB(r)=19$ mag and $AB({\rm
  F814W})=18.66$ mag.  The close proximity to a bright galaxy may have
made object deblending challenging in ground-based images.  As
discussed below, the MUSE spectrum of this galaxy exhibits absorption
features that are consistent with a passive galaxy at $z=0.3303$,
similar to the LRG but nearly four magnitudes fainter.  We therefore
estimated the $g$-, $r$-, and $z$-band magnitudes of this faint
satellite based on the observed brightness of the LRG in these
bandpasses.

For every object detected in the F814W image, we extracted a mean
spectrum and the associated error spectrum over a circular aperture
centered at the object position in the MUSE data cube.  The aperture
size was adjusted according to the object brightness.  Within the MUSE
field, two objects (the QSO and LRG) have optical spectra available
from the SDSS spectroscopic archive.  Comparisons of SDSS spectra and
our own MUSE spectra of these objects show good agreement both in the
locations of the line features and in the relative fluxes across the
full spectral range from $\lambda\approx 4900$ \AA\ to $\lambda>9200$
\AA, confirming the accuracy of wavelength and flux
calibrations in the MUSE data cube.

\subsection{LDSS3 multi-object spectroscopic observations}

Multi-object spectroscopic observations of the QSO field were obtained
using the Low Dispersion Survey Spectrograph 3 (LDSS3) and the VPH-all
grism on the Magellan Clay telescope.  LDSS3 observes a field of $4'$
in radius at a pixel scale of $0.189''$, expanding the redshift survey
beyond the MUSE field of view and providing a wider view of the
large-scale galactic environment around the LRG and the pristine LLS
at $z=0.33$.  The VPH-all grism covers a wavelength range of
$4500-10000$ \AA\ with a spectral resolution of ${\rm FWHM}\approx
400$ \kms.  Multiple slitmasks were designed to observe a total of 247
objects with $AB(r)<23.5$ mag which, at $z=0.33$, corresponds to
$>0.05\,L_*$ in intrinsic luminosity.  Objects at $<30''$ in angular
distance from the QSO sightline were excluded from the LDSS3
observations to minimize duplicated spectroscopic follow-up effort,
given the existing MUSE data.  The observations were carried out in
March 2018 and each mask was exposed for a total of 80 minutes under
sub-arcsecond seeing conditions.  The data were reduced using the
COSMOS pipeline described in Chen \& Mulchaey (2009).  

\subsection{The redshift survey}

To determine the redshift of each object, we performed a $\chi^2$
fitting routine that compares the observed spectrum with models formed
from a linear combination of four eigenspectra from the SDSS at
different redshifts (see Chen \& Mulchaey 2009 and Johnson \etal\ 2013
for a detailed description).  The best-fit redshift of each object
returned from the $\chi^2$ routine was visually inspected for
confirmation.  

From the MUSE data, we were able to obtain robust redshift
measurements for 67 F814W-identified objects.  Redshift uncertainties
based on MUSE spectra are typically $\Delta\,z\apl \pm 0.0001$ for
emission-line galaxies and $\Delta\,z\approx \pm 0.0003$ for
absorption-line galaxies.  All but one of the 42 objects brighter than
$AB({\rm F814W})=24$ mag in the $1'\times 1'$ MUSE field are
identified with a spectroscopic redshift, reaching nearly 100\% survey
completeness for objects with $AB({\rm F814W})<24$ mag.  In addition,
26 galaxies fainter than $AB({\rm F814W})=24$ mag have redshift
identifications based on observations of strong emission lines,
maintaining a 50\% completeness level for objects with $AB({\rm
  F814W})=24-25$ mag.  The faintest spectroscopically identified
galaxy is J135726.81$+$043544.27 of $AB({\rm F814W})\approx 27$ mag at
$z=0.3313$, close in redshift and projected distance to the LRG and
the LLS.

From the LDSS3 observations, we were able to obtain robust redshift
measurements for 146 objects of $AB(r)<23.4$ mag at angular
separations of $0.5'- 3.3'$ from the QSO sightline, including six
objects that were also covered by the MUSE observations.  Redshift
uncertainties based on LDSS3 spectra are typically $\Delta\,z\approx
\pm 0.0003$.  Over the annulus of $0.5'-2'$ radius of the QSO
sightline, corresponding to $150-570$ pkpc at $z=0.33$, the LDSS3
survey reached $>80$\% completeness for extended sources of $AB(r)\le
22$ mag, $>70$\% for objects of $AB(r)<23$ mag, and $>60$\% for
objects of $AB(r)<23.5$ mag.

The results of the LDSS3 survey are summarized in Figure 2 with each
spectroscopically identified object marked by the best-fit redshift.
The redshift measurements are shown to two decimal places to avoid
crowding.  Galaxies found near the LLS from the combined LDSS3 and
MUSE survey are labeled G1 through G12 in red.  Redshift measurements
of these galaxies are presented in column (4) of Table 1.  Figure 2
shows that the LDSS3 and MUSE survey data together have uncovered an
overdensity of galaxies near the LLS within 500 pkpc in radius.

The results of the MUSE survey in the inner $1'\times 1'$ field around
the QSO are summarized in Figure 3, which demonstrates that the
high-throughput and dense sampling of MUSE enable a high survey
completeness to faint magnitudes (see \S\ 3 below).  The left panel
shows the {\it HST}/ACS F814W image, while the right panel displays a
narrow-band image of 20\,\AA\ wide, centered at 6660 \AA\ from the
combined MUSE data cube.  The wavelength range is chosen to match the
expected wavelength of \OIII\,$\lambda\,5007$ at the redshift of this
galaxy group, $z=0.33$.  The circles highlight the three line-emitting
objects that are barely detected in the F814W band.


\begin{figure*}
\begin{adjustbox}{
addcode= {\begin{minipage}{\width}} {\caption{Zoom-in view of the
    field centered on QSO SDSSJ\,135726.27$+$043541.4 at $z_{\rm
      QSO}=1.233$, with the {\it HST}/ACS F814W image displayed on the
    {\it left} and a narrow-band image of 20 \AA\ wide, centered at
    6660 \AA\ from the combined MUSE data cube displayed on the {\it
      right}.  North is up and east to the left.  At $z=0.33$, the
    wavelength range of the narrow-band image corresponds to the
    observed wavelength of \OIII\,$\lambda\,5007$.  The dashed circle
    marks the $30''$ radius from the QSO sightline, corresponding to
    $d\approx 140$ pkpc at $z=0.33$.  Similar to Figure 2,
    spectroscopically identified objects are marked with their
    redshift measurements and galaxies associated with the pristine
    LLS at $z=0.33$ are highlighted in red.  In the MUSE narrow-band
    image, precise redshifts of the galaxies in the vicinity of the
    LLS (G1 through G7) are presented.  The F814W image and MUSE data
    together provide a detailed view of additional neighbors closer in
    projected distance to the LLS than the LRG (G5), three of which
    are emission-line dominated galaxies with exceedingly faint
    continuum fluxes (G1, G2, and G3; also marked in circles).
  }\label{figure:images}\end{minipage}}, rotate=90,center}
  \includegraphics[scale=0.65]{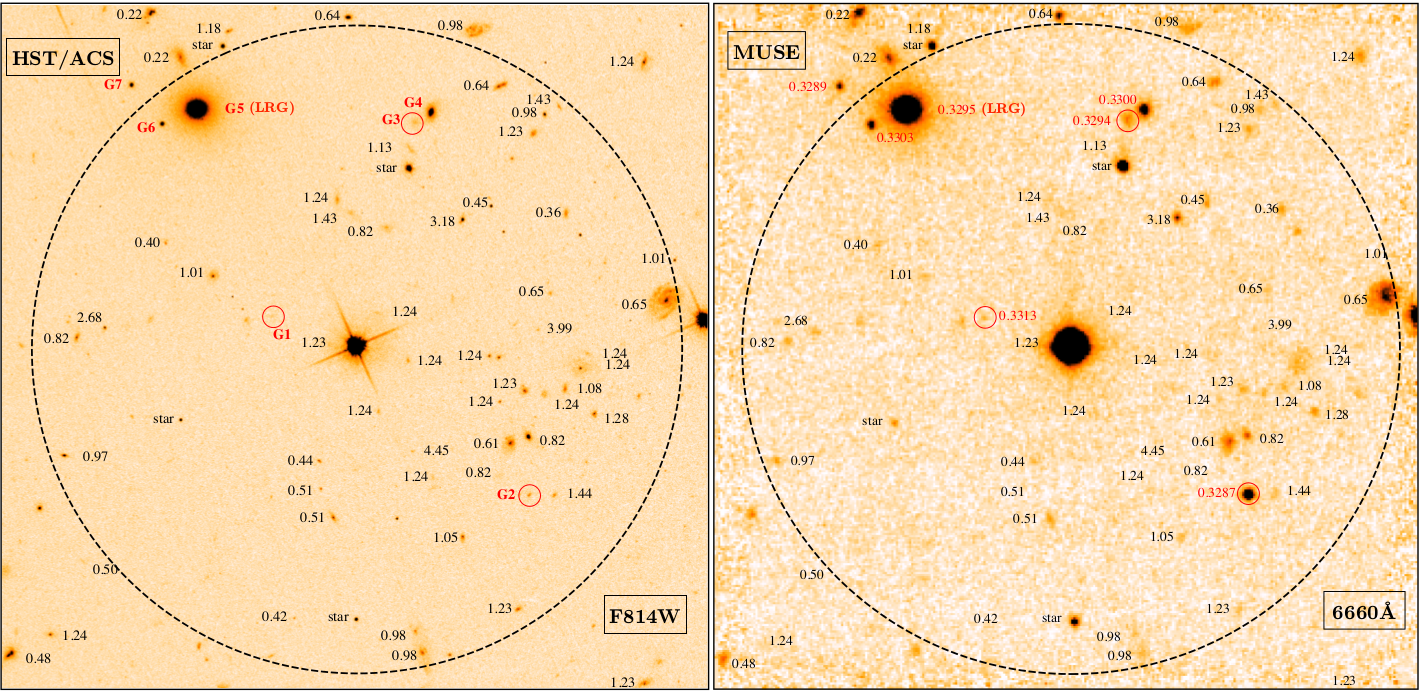}
\end{adjustbox}

\end{figure*}


\begin{table*}
\small
\centering
\caption{Photometric properties of galaxies in the vicinity of the LLS at $z=0.3287$}
\label{table:sample}
\centering {
\begin{tabular}{llrcrcccc}
\hline \hline
  &     & \multicolumn{1}{c}{$\theta$} &              & \multicolumn{1}{c}{$d$} & & & & \\
\multicolumn{1}{c}{ID} & \multicolumn{1}{c}{Coordinates} & \multicolumn{1}{c}{($''$)}  & $z$ & \multicolumn{1}{c}{(pkpc)} & \multicolumn{1}{c}{$AB(g)$} & \multicolumn{1}{c}{$AB(r)$}& \multicolumn{1}{c}{$AB(z)$}& \multicolumn{1}{c}{$AB({\rm F814W})$} \\
\multicolumn{1}{c}{(1)} & \multicolumn{1}{c}{(2)}  & \multicolumn{1}{c}{(3)}  & \multicolumn{1}{c}{(4)} & \multicolumn{1}{c}{(5)} & \multicolumn{1}{c}{(6)}& \multicolumn{1}{c}{(7)}& \multicolumn{1}{c}{(8)}& \multicolumn{1}{c}{(9)} \\
\hline
G1 & J135726.81$+$043544.27 & 8.2 & 0.3313 & 39 & $>25.0$ & $>24.3$ & $>23.6$ & 26.89$\pm$0.17 \\
G2 & J135725.20$+$043527.99 & 21.2 & 0.3287 & 100 & 25.44$\pm$0.21 & 24.79$\pm$0.18 & 25.09$\pm$0.63 & 24.74$\pm$0.04 \\
G3 & J135725.91$+$043602.48 & 21.6 & 0.3294 & 102 & $>25.0$ & $>24.3$ & $>23.6$ & 23.80$\pm$0.03 \\
G4 & J135725.81$+$043603.42 & 22.9 & 0.3300 & 109 & 23.15$\pm$0.04 & 21.69$\pm$0.02 & 20.88$\pm$0.02 & 21.31$\pm$0.01 \\
G5$^a$ & J135727.28$+$043603.38 & 26.5 & 0.3295 & 126 & 20.61$\pm$0.01 & 18.95$\pm$0.01 & 18.07$\pm$0.01 & 18.60$\pm$0.01 \\
G6$^b$ & J135727.50$+$043601.98 & 27.4 & 0.3303 & 130 & $24.3$ & $22.7$ & $21.8$ & 22.29$\pm$0.01 \\
G7 & J135727.69$+$043605.55 & 32.0 & 0.3289 & 152 & 24.61$\pm$0.11 & 23.03$\pm$0.05 & 22.24$\pm$0.04 & 22.61$\pm$0.01 \\
G8 & J135729.10$+$043539.23 & 42.1 & 0.3268 & 199 & 22.85$\pm$0.04 & 22.33$\pm$0.04 & 22.10$\pm$0.06 & 22.31$\pm$0.01 \\
G9 & J135723.01$+$043510.54 & 58.1 & 0.3292 & 276 & 23.45$\pm$0.05 & 22.23$\pm$0.03 & 21.49$\pm$0.04 & 21.83$\pm$0.01 \\
G10 & J135723.27$+$043630.66 & 66.6 & 0.3293 & 316 & 23.68$\pm$0.06 & 22.64$\pm$0.04& 22.08$\pm$0.06 & 22.22$\pm$0.01 \\
G11 & J135727.09$+$043655.27 & 74.6 & 0.3294 & 354 & 23.38$\pm$0.06 & 22.23$\pm$0.04 & 21.50$\pm$0.04 & 21.81$\pm$0.01 \\
G12 & J135730.69$+$043629.38 & 81.3 & 0.3290 & 386 & 23.74$\pm$0.06 & 22.35$\pm$0.04 & 21.60$\pm$0.04 & 21.97$\pm$0.01 \\
\hline
\multicolumn{9}{l}{$^\mathrm{a}$The LRG identified in the SDSS spectroscopic sample.  The redshift determined from MUSE data agrees} \\
\multicolumn{9}{l}{\ \ very well with the SDSS measurement.} \\
\multicolumn{9}{l}{$^\mathrm{b}$This object is not identified in the DECaLS catalog, possibly due to blending with the LRG at $\approx 3''$ away.} \\
\multicolumn{9}{l}{\ \ We estimate the $g,r,z$ magnitudes based on its observed flux ratio relative to the LRG in the F814W band.} \\
\end{tabular}
}
\end{table*}

\section{Galactic Environment of the LLS at $z = 0.33$}

The analysis presented in \S\ 2 yielded a highly complete
spectroscopic sample of galaxies in the field around QSO
SDSSJ\,135726.27$+$043541.4.  Specifically at angular distances
$\theta\apl 30''$ from the sightline, the MUSE survey reached nearly
100\% completeness for objects with $AB({\rm F814W})<24$ mag and 50\%
completeness for objects with $AB({\rm F814W})=24-25$ mag.  At
$z=0.33$, the observed F814W bandpass corresponds well with rest-frame
$r$-band, and from Cool \etal\ (2012) blue galaxies at this redshift
are found to have a characteristic rest-frame absolute $r$-band
magnitude of $M_{r_*}=-21.3$ and red galaxies have $M_{r_*}=-21.5$.
An observed brightness of $AB({\rm F814W})=24$ (25) mag therefore
corresponds to $\approx 0.02 (0.007)\,L_*$ at $z=0.33$.  

Over the angular distance range from $\theta=30''$ to $\theta=2'$, the
LDSS3 survey reached $>80$\% completeness for objects with $AB(r)<22$
mag and $>70$\% completeness for objects with $AB(r)<23$ mag.  At
$z=0.33$, the survey limit at $AB(r)=23$ mag corresponds to
$0.08\,L_*$ for quiescent galaxies and $0.05\,L_*$ for star-forming
galaxies.  The highly complete imaging and spectroscopic data
therefore provide an ultra deep view of the galactic environment of
the chemically-pristine LLS and the nearby LRG.

Indeed, our imaging and spectroscopic survey of this field has
uncovered a group of 12 galaxies in the vicinity of the LLS at
$z=0.33$.  The photometric properties of these galaxies are presented
in columns (6) through (9) of Table 1, in increasing angular distance
from the QSO ($\theta$ in column 3).  For objects that are not
detected in the DECaLS images, the 5-$\sigma$ limiting magnitudes
computed for extended sources in this region by the DECaLS team (e.g.,
Dey \etal\ 2018) are provided in Table 1.


In addition to the previously identified LRG at $d=126$ pkpc, our
survey yielded 11 new galaxies with rest-frame absolute $r$-band
magnitude ranging from $M_{r}=-14$ to $M_{r}=-19.6$ (corresponding to
a luminosity range of $0.001\,L_* - 0.2\,L_*$) at $d=39-386$ pkpc, and
no additional member at $d>400$ pkpc.  In comparison to the LRG, a
$2\,L_*$ evolved galaxy (estimated based on the observed F814W
brightness and $\approx 30$\% less luminous than the previous estimate
from SDSS) with an implied virial radius of $\approx 500$ pkpc (e.g.,
Gauthier \etal\ 2009; Huang \etal\ 2016), the newly identified
galaxies are all significantly fainter and the three galaxies closest
to the QSO sightline are line emitters with exceedingly faint optical
continuum magnitudes, $AB(r)>24.4$ mag (galaxies G1, G2, \& G3 circled
in Figure 3).


We infer a stellar mass, \mstar, for each member of the galaxy `group'
based on $M_r$ and the rest-frame $g-r$ color following the
prescription provided in Johnson \etal\ (2015).  The relation is
calibrated for a Chabrier (2003) initial mass function with typical
uncertainties in \mstar\ of less than 0.15 dex (see Johnson
\etal\ 2015 for a detailed discussion).  While the three
line-emitters, G1, G2, \& G3, are not detected in the DECaLS images,
the MUSE spectra display blue continua.  We adopt the blue branch from
Johnson \etal\ (2015), when inferring \mstar\ for these galaxies.  For
the remaining member galaxies, we estimate \mstar\ based on the
rest-frame $g-r$ color computed from available DECaLS photometry.
Independent \mstar\ estimates using the FAST code (Kriek \etal\ 2009)
are also available for objects observed in the MUSE frame (see Segers
\etal\ 2018).  Comparisons between our \mstar\ estimates and those
from FAST show good agreements to within measurement uncertainties for
all but one object.  Galaxy G3 is the only exception with a
FAST-estimated \mstar\ 0.6 dex higher than ours.  But because G3 is
merely $1.8''$ from G4, object blending in MUSE data may impose
additional systematic uncertainties in the \mstar\ measurement.  In
what follows, we describe the properties of individual group members.


\begin{table*}
\small
\centering
\caption{Intrinsic properties of galaxies in the vicinity of the LLS at $z=0.3287$}
\label{table:group}
\centering {
\begin{tabular}{lrrrcrcrrr}
\hline \hline
                           &                          & \multicolumn{1}{c}{$d$}   & \multicolumn{1}{c}{$\Delta\,v$} &                           &                          &                         & \multicolumn{1}{c}{EW(H$\alpha$)$^b$} &   \multicolumn{1}{c}{SFR$^c$}    &   \multicolumn{1}{c}{ISM} \\
\multicolumn{1}{c}{Galaxy} & \multicolumn{1}{c}{$z$}    & \multicolumn{1}{c}{(pkpc)} & \multicolumn{1}{c}{(\kms)}       & \multicolumn{1}{c}{$M_r$$^a$} & \multicolumn{1}{c}{$\log\,M_{\rm star}/{\rm M}_\odot$} & \multicolumn{1}{c}{Type} & \multicolumn{1}{c}{(\AA)} & \multicolumn{1}{c}{(${\rm M}_\odot\,{\rm yr}^{-1}$)} & \multicolumn{1}{c}{(O/H)$^d$} \\
\multicolumn{1}{c}{(1)}    & \multicolumn{1}{c}{(2)}  & \multicolumn{1}{c}{(3)}   & \multicolumn{1}{c}{(4)}         & \multicolumn{1}{c}{(5)}   & \multicolumn{1}{c}{(6)}  & \multicolumn{1}{c}{(7)} & \multicolumn{1}{c}{(8)} & \multicolumn{1}{c}{(9)} & \multicolumn{1}{c}{(10)} \\
\hline
G1  & 0.3313 &  39 & $+$587 &  $-14.0$ &  6.9 &   SF    & $-59\pm 10$     & 0.005   & $\sim 7.5^e$          \\
G2  & 0.3287 & 100 &      0 &  $-16.2$ &  7.8 &   SF    & $-121.5\pm 2.6$ & 0.10    & $7.8\pm 0.1$  \\
G3  & 0.3294 & 102 & $+$158 &  $-17.1$ &  8.1 &   SF    & $-23.2\pm 1.2$  & 0.03    & $7.5\pm 0.1$  \\
G4  & 0.3300 & 109 & $+$294 &  $-19.6$ & 10.0 & passive & $4.0\pm 0.6$    & $<0.11$ & ... \\ 
G5  & 0.3295 & 126 & $+$181 &  $-22.3$ & 11.1 & passive & $2.7\pm 0.2$    & $< 0.9$ & ... \\ 
G6  & 0.3303 & 130 & $+$361 &  $-18.6$ &  9.6 & passive & $4.3\pm 1.0$    & $<0.05$ & ... \\ 
G7  & 0.3289 & 152 &  $+$45 &  $-18.3$ &  9.5 & passive & $5.2\pm 1.0$    & $<0.04$ & ... \\ 
G8  & 0.3268 & 199 & $-$429 &  $-18.6$ &  8.7 &   SF    & $-84.6\pm 3.3$  & 0.30    &  $8.3\pm 0.1$ \\
G9  & 0.3292 & 276 & $+$112 &  $-19.1$ &  9.7 & passive & $-4.0\pm 0.6$   & $<0.005$ &  ... \\
G10 & 0.3293 & 316 & $+$135 &  $-18.7$ &  9.3 & post SB &  $-10.2\pm 1.0$ & 0.04 & ... \\
G11 & 0.3294 & 354 & $+$158 &  $-19.1$ &  9.7 & passive & $-11.7\pm 1.2$  & $<0.06$ &  ... \\
G12 & 0.3290 & 386 &  $+$68 &  $-19.0$ &  9.6 & passive & $<2$  & $<0.01$ &  ... \\
\hline

\multicolumn{10}{l}{$^\mathrm{a}$At $z=0.3$, star-forming galaxies have a characteristic rest-frame absolute $r$-band magnitude of $M_{r_*}=-21.3$ and passive galaxies} \\
\multicolumn{10}{l}{\ \ have $M_{r_*}=-21.5$ (e.g., Cool \etal\ 2012).} \\
\multicolumn{10}{l}{$^\mathrm{b}$Rest-frame H$\alpha$ equivalent width.  Negative values represent emission, while positive values represent absorption.} \\
\multicolumn{10}{l}{$^\mathrm{c}$For star-forming (SF) galaxies, SFR is estimated based on $L({\rm H}\alpha)$, while for passive galaxies a 2-$\sigma$ upper limit is }\\
\multicolumn{10}{l}{\ \  estimated from the non-detection of [O\,II]\,$\lambda\lambda\,3726, 3729$.} \\
\multicolumn{10}{l}{$^\mathrm{d}$For comparison, the oxygen abundance of the Sun is $12 + \log({\rm O}/{\rm H}) = 8.69\pm 0.05$ (Asplund \etal\ 2009).} \\
\multicolumn{10}{l}{$^\mathrm{e}$Based on the mass-metallicity relation of local dwarf galaxies from Berg \etal\ (2012) with a dispersion of $\sigma_{\rm (O/H)}=0.15$ dex.} \\
\end{tabular}
}
\end{table*}

G1 is an ultra faint dwarf galaxy found at $\theta=8.2''$ and
$z=0.3313$, with corresponding projected distance $d=39$ pkpc and
line-of-sight velocity $\Delta\,v=+587$ \kms\ from the LLS.  The
galaxy has $M_r=-14$, corresponding to $0.001\,L_*$, and
$\log\,\mstar/\msun=6.9$.  The optical spectrum of this galaxy is
dominated by emission-line features, typical of a star-forming galaxy.
We measure a rest-frame H$\alpha$ equivalent width of $\ewha=-59\pm
10$ \AA, a total line flux of $f({\rm H}\alpha)=(2.8\pm 0.4)\times
10^{-18}\,{\rm erg}\,{\rm s}^{-1}\,{\rm cm}^{-2}$, and a luminosity
of $L({\rm H}\alpha)=(9.7\pm 1.4)\times 10^{38}\,{\rm erg}\,{\rm
  s}^{-1}$ which, following the calibration of Kennicutt \& Evans
(2012), leads to a star formation rate (SFR) of $(5\pm 0.8)\times
10^{-3}$\,\sfr.  To estimate the ISM gas phase metallicity, we
consider both the $N2\equiv\log\,{\rm [N\,II]}\,\lambda\,6585/{\rm
  H}\alpha$ index and the $R_{23}\equiv\,({\rm
  [O\,II]}\,\lambda\lambda\,3726,3729+{\rm
  [O\,III]}\,\lambda\lambda\,4959,5007)/{\rm H}\beta$ index.  However,
the expected location for the [N\,II]\,$\lambda\,6585$ line is
contaminated by strong OH lines.  In addition, while the ${\rm
  [O\,III]}\,\lambda\lambda\,4959,5007$ lines are detected at a high
confidence level, the spectrum has insufficient quality for detecting
H$\beta$ and [O\,II]\,$\lambda\lambda\,3726, 3729$.  Although we are
unable to obtain a direct measurement of the gas-phase metallicity in
the interstellar medium (ISM) of this object, we infer a gas-phase
metallicity of $12 + \log({\rm O}/{\rm H}) = 7.5\pm 0.2$ based on its
\mstar\ and the best-fit mass-metallicity relation of low-mass dwarf
galaxies from Berg \etal\ (2012).

G2 is a faint dwarf galaxy found at $\theta=21.2''$ and $z=0.3287$,
with corresponding projected distance $d=100$ pkpc and line-of-sight
velocity $\Delta\,v=0$ \kms\ from the LLS.  The galaxy has
$M_r=-16.1$, corresponding to $0.008\,L_*$, and
$\log\,\mstar/\msun=7.7$.  The optical spectrum of this galaxy is
dominated by emission-line features, typical of a star-forming galaxy.
We measure a rest-frame H$\alpha$ equivalent width of $\ewha=-121.5\pm
2.6$ \AA, a total H$\alpha$ line flux of $f({\rm H}\alpha)=(5.1\pm
0.1)\times 10^{-17}\,{\rm erg}\,{\rm s}^{-1}\,{\rm cm}^{-2}$, and a
luminosity of $L({\rm H}\alpha)=(1.8\pm 0.03)\times 10^{40}\,{\rm
  erg}\,{\rm s}^{-1}$, which leads to ${\rm SFR}= 0.097\pm
0.002\,\sfr$.  Based on the non-detection of [N\,II]\,$\lambda\,6585$
and the calibration from Marino \etal\ (2013), $12 + \log({\rm O}/{\rm
  H}) = 8.74 + 0.46\times N2$, we infer a 2-$\sigma$ upper limit to
the ISM gas-phase metallicity of $12 + \log({\rm O}/{\rm H}) < 8.1$.
The observed H$\alpha$/H$\beta$ ratio indicates that dust extinction
is negligible in this galaxy.  We therefore estimate the oxygen
abundance from $R_{23}$, assuming no extinction corrections.  Based on
the calibration of Yin \etal\ (2007), $12 + \log({\rm O}/{\rm H}) =
6.486 + 1.401\times \log\,R_{23}$, we find $12 + \log({\rm O}/{\rm H})
= 7.8\pm 0.1$ for the observed flux ratio of $R_{23}=8.9\pm 0.3$ with
the error in gas metallicity dominated by the uncertainty in the
$R_{23}$ calibration.

G3 is a faint dwarf galaxy found at $\theta=21.6''$ and $z=0.3294$,
with corresponding projected distance $d=102$ pkpc and line-of-sight
velocity $\Delta\,v=+158$ \kms\ from the LLS.  The galaxy has
$M_r=-17$, corresponding to $0.02\,L_*$, and $\log\,\mstar/\msun=8.1$.
The optical spectrum of this galaxy is dominated by emission-line
features, typical of a star-forming galaxy.  We measure a rest-frame
H$\alpha$ equivalent width of $\ewha=-23.2\pm 1.2$ \AA, a total
H$\alpha$ line flux of $f({\rm H}\alpha)=(1.7\pm 0.1)\times
10^{-17}\,{\rm erg}\,{\rm s}^{-1}\,{\rm cm}^{-2}$, and a luminosity
of $L({\rm H}\alpha)=(5.9\pm 0.3)\times 10^{39}\,{\rm erg}\,{\rm
  s}^{-1}$, which leads to ${\rm SFR}= 0.032\pm 0.002\,\sfr$.  Based
on the non-detection of [N\,II]\,$\lambda\,6585$, we infer a
2-$\sigma$ upper limit to the ISM gas-phase metallicity of $12 +
\log({\rm O}/{\rm H}) < 8.2$.  Applying no extinction corrections
based on the observed H$\alpha$/H$\beta$ ratio, we find based on the
$R_{23}$ index an oxygen abundance of $12 + \log({\rm O}/{\rm H}) =
7.5\pm 0.1$ for the observed flux ratio of $R_{23}=5.9\pm 0.8$.

Galaxies G4, G5 (the LRG), G6, and G7 at $d=109-152$ pkpc and
$\Delta\,v = 45 - 361$ \kms\ all exhibit strong absorption features
such as Ca\,II H\&K, G-band, Mg\,I triplet, and Balmer transitions,
with no trace of emission lines.  These galaxies have $M_r=-18.6 -
-22.3$, corresponding to $0.07-2\,L_*$, and $\log\,\mstar/\msun\approx
9.6-11.1$.  The rest-frame H$\alpha$ equivalent width ranges from
$\ewha=2.7$ \AA\ to 5.2 \AA.  The observed spectral features confirm
that these are evolved galaxies with little/no ongoing star formation.
We infer 2-$\sigma$ upper limits on the underlying SFR based on the
absence of [O\,II] emission and the star formation calibrator of
Kewley \etal\ (2004), assuming that no dust extinction correction is
necessary for these quiescent galaxies.
To gauge the chemical enrichment in these early-type galaxies, we
estimate their stellar metallicities based on the Fe\,I line at 4668
\AA\ and H$\gamma$ that are known to be sensitive to the underlying
stellar age and metallicity.  Following the definitions of these
indices from Kuntschner \& Davies (1998), we find that the C4668
equivalent width of these four galaxies ranges from 4.2 \AA\ to 6.3
\AA, and the H$\gamma$ equivalent width ranges from $-6.8$ \AA\ to
$-3.0$ \AA, leading to a mean stellar metallicity that exceeds solar
and a mean stellar age of $\apg 2$ Gyr for these galaxies (see Figure
2 of Kuntschner \& Davies 1998).

G8 is a star-forming dwarf galaxy found at $\theta=42.1''$ and
$z=0.3268$, with corresponding projected distance $d=199$ pkpc and
line-of-sight velocity $\Delta\,v=-429$ \kms\ from the LLS.  The
galaxy has $M_r=-18.6$, corresponding to $0.09\,L_*$, and
$\log\,\mstar/\msun=8.7$.  The optical spectrum of this galaxy
exhibits strong emission-line features, typical of a star-forming
galaxy.  We measure a rest-frame H$\alpha$ equivalent width of
$\ewha=-84.6\pm 3.3$ \AA, a total H$\alpha$ line flux of $f({\rm
  H}\alpha)=(1.69\pm 0.07)\times 10^{-16}\,{\rm erg}\,{\rm
  s}^{-1}\,{\rm cm}^{-2}$, and a luminosity of $L({\rm
  H}\alpha)=(5.5\pm 0.2)\times 10^{40}\,{\rm erg}\,{\rm s}^{-1}$,
which leads to ${\rm SFR}= 0.30\pm 0.01\,\sfr$.  Based on the observed
[N\,II]\,$\lambda\,6585$/H$\alpha$\,$\lambda\,6564$ ratio, we infer an
ISM gas-phase metallicity of $12 + \log({\rm O}/{\rm H}) = 8.3\pm
0.1$, $\approx 50$\% solar.

\begin{figure*}
\includegraphics[scale=0.95]{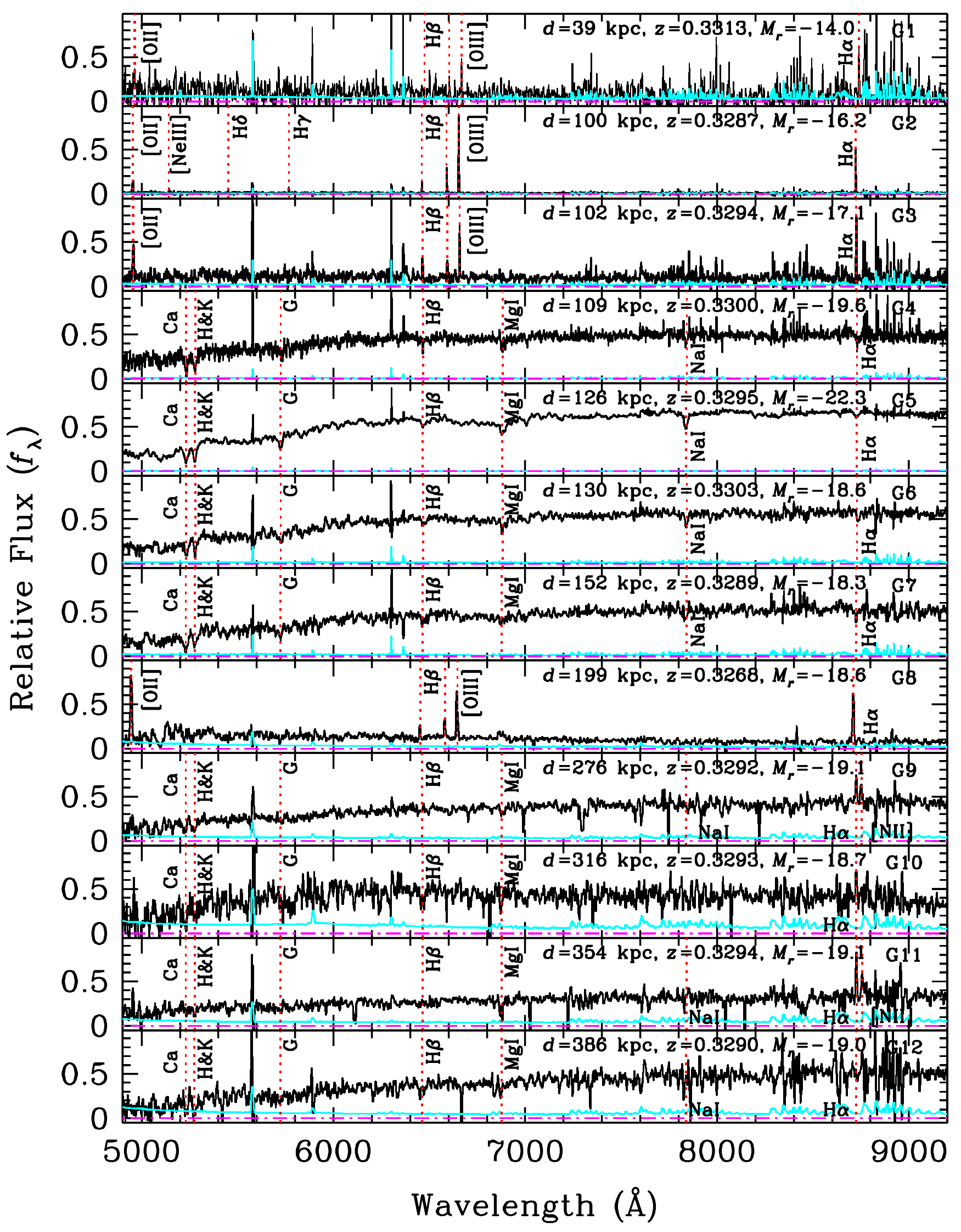}
\caption{Optical spectra of 12 galaxies found in the vicinity of the
  chemically-pristine Lyman limit absorber at $z=0.3287$ (Papers
  I\&II) with increasing projected distance from top to bottom.  The
  corresponding 1-$\sigma$ error spectra are shown in cyan.  Prominent
  spectral features are highlighted in red dotted lines, marked with
  the corresponding line identifications.  The best-fit redshift and
  the corresponding projected distance ($d$) and rest-frame absolute
  $r$-band magnitude ($M_r$) are listed on the right.  The spectral
  panels are organized with increasing projected distances from top to
  bottom.  Galaxy spectra of G1 through G7 are from MUSE, while the
  remaining five are from the LDSS3 observations.  All three galaxies
  at $d\apl 100$ pkpc are emission-line dominated faint galaxies
  ($<0.02\,L_*$), while the LRG (G5) remains to be the most dominant
  galaxy (contributing $\approx 80$\% in the total mass) in this
  galaxy `group'.} \label{figure:spectra}
\end{figure*}

Galaxies G9 and G11 at $d=276$ and 354 pkpc and $\Delta\,v = +112$ and
$+158$ \kms\ both exhibit strong absorption features such as Ca\,II
H\&K, G-band, Mg\,I triplet, typical of a quiescent galaxy.  At the
same time, the spectra also displays H$\alpha$ and strong
[N\,II]\,$\lambda\,6585$ in emission, resembling nearby LINER-like
galaxies (e.g., Sarzi \etal\ 2006; Yan \etal\ 2006).  We measure a
rest-frame H$\alpha$ equivalent width of $\ewha=-4.0\pm 0.6$ \AA\ and
${\rm EW}({\rm [N\,II]}\lambda\,6585)=-3.4\pm 0.7$ \AA\ for G9 and
$\ewha=-11.7\pm 1.2$ \AA\ and ${\rm EW}({\rm
  [N\,II]}\lambda\,6585)=-7.1\pm 1.9$ \AA\ for G11.  These two
galaxies have $M_r\approx -19.1$, corresponding to $0.1\,L_*$, and
$\log\,\mstar/\msun\approx 9.7$.  Given the likely presence of
additional ionizing sources that generated the emission features
observed in the galaxy, we infer an upper limit on the underlying SFR
of ${\rm SFR}<0.02\,\sfr$ for G9 and ${\rm SFR}<0.06\,\sfr$ for G11
based on the observed H$\alpha$ line flux.

G10 is a post-starburst galaxy at $\theta=66.6''$ and $z=0.3293$, with
corresponding projected distance $d=316$ pkpc and line-of-sight
velocity $\Delta\,v=+135$ \kms\ from the LLS.  The optical spectrum of
this galaxy exhibits a combination of strong Ca\,II H\&K and the
Balmer absorption series, and weak emission features.  The galaxy has
$M_r=-18.7$, corresponding to $0.09\,L_*$, and
$\log\,\mstar/\msun=9.3$.  We measure a rest-frame H$\alpha$
equivalent width of $\ewha=-10.2\pm 1.0$ \AA\ after correcting for
stellar absorption, and find a corresponding total H$\alpha$ line flux
of $f({\rm H}\alpha)=(2.2\pm 0.2)\times 10^{-17}\,{\rm erg}\,{\rm
  s}^{-1}\,{\rm cm}^{-2}$, and a luminosity of $L({\rm
  H}\alpha)=(7.2\pm 0.7)\times 10^{39}\,{\rm erg}\,{\rm s}^{-1}$,
which leads to ${\rm SFR}= 0.04\pm 0.004\,\sfr$.  The spectrum does
not have sufficient quality for placing a sensitive constraint for the
associated [N\,II] lines.  Based on the non-detection of
[N\,II]\,$\lambda\,6585$, we infer a 2-$\sigma$ upper limit to the ISM
gas-phase metallicity of $12 + \log({\rm O}/{\rm H}) < 8.6$.

G12 is another quiescent galaxy at $d=386$ pkpc and $\Delta\,v = +68$
\kms\ with strong absorption features such as Ca\,II H\&K, G-band, and
Mg\,I triplet and no trace of emission lines.  This galaxy has
$M_r\approx -19$, corresponding to $0.1\,L_*$, and
$\log\,\mstar/\msun\approx 9.6$.  We infer an upper limit on the
underlying SFR of ${\rm SFR}<0.01\,\sfr$ based on the absence of
H$\alpha$ line.

We summarize the properties of all 12 galaxies in Table 2, which lists
the redshift $z$, projected distance $d$, line-of-sight velocity
offset from the LLS $\Delta\,v$, rest-frame $r$-band magnitude $M_r$,
stellar mass $\mstar$, galaxy type (star-forming versus passive),
rest-frame H$\alpha$ equivalent width $\ewha$, star formation rate
SFR, and ISM gas-phase metallicity in columns (2) through (10).  The
extracted optical spectra of galaxies 
are also presented in Figure 4, along with the 1-$\sigma$ error
spectra.

\section{Discussion and Conclusions}

The analysis presented in \S\ 3 demonstrates that combining deep {\it
  HST} images, MUSE observations, and wide-field survey data enables
an unprecedented, ultra-deep view of the galactic environment of a
chemically-pristine LLS at $z=0.33$.  Previous efforts in
characterizing the galactic environment of strong \HI\ absorbers at
$z<1$ include the identification of a gas-rich galaxy group near a
damped \lya\ absorber (DLA) of $\log\,N({\rm HI})/{\rm
  cm}^{-2}=21.7\pm 0.1$ and $\approx 12$\% solar metallicity at
$z=0.313$ (Chen \& Lanzetta 2003; Kacprzak \etal\ 2010), the discovery
of a metal-poor (${\rm [O/H]}=-1.6\pm 0.1$) strong LLS of
$\log\,N({\rm HI})/{\rm cm}^{-2}=19.3$ at $z_{\rm LLS}=0.0063$ in the
outskirts of the Virgo cluster (Tripp \etal\ 2005), and a star-forming
galaxy at $d=54$ kpc from a strong LLS of $\log\,N({\rm HI})/{\rm
  cm}^{-2}=19.1$ and $\apg 10$\% solar metallicity at $z=0.78$
(Rahmani \etal\ 2018).  In the first case, the metallicity of the DLA
was found to be similar to the ISM metallicity in some of the group
members, indicating that the DLA originates in tidal debris in the
group environment.  In the second case, the LLS was found to be
associated with the NGC\,4261 group with the nearest galaxy being a
star-forming galaxy of $0.25\,L_*$ at $d=86$ kpc from the absorber,
and the metal-poor LLS was attributed to the ISM removed from dwarf
satellites (however, see Chengalur \etal\ 2015 for more discussions).
In the third case, the LLS was moderately enriched and was attributed
to galactic winds from the disk galaxy.  In all cases, the galaxy
survey data are too incomplete to place stringent constraints for the
possible presence of ultra-faint dwarf satellites.

Our spectroscopic survey in the inner 150 pkpc radius of the QSO
sightline reaches a 100\% completeness level for galaxies of
luminosity $>0.02\,L_*$ at the redshift of the absorber, and maintains
a 50\% completeness level for galaxies as faint as $0.007\,L_*$.
Beyond $d=150$ kpc, the LDSS3 survey reaches $>70$\% completeness for
galaxies of $>0.05\,L_*$.  A group of 12 galaxies are found at $d\apl
400$ pkpc and $|\Delta\,v|<600$ \kms\ of the optically-thick absorber,
with intrinsic luminosity ranging from $0.001\,L_*$ to $2\,L_*$ and a
corresponding stellar mass range of $\mstar\approx 10^{7-11}\,\msun$.
The wide-area LDSS3 survey covered an area out to $d\approx 1$ Mpc in
projected distance from the QSO sightline, but it did not uncover
additional galaxies associated with the galaxy `group' at $d>400$
pkpc.

The group of 12 galaxies in the vicinity of the LLS contains a total
stellar mass of $\mstar({\rm total})=1.6\times 10^{11}\,\msun$,
$\approx 80$\% of which resides in the LRG.  Considering all 12
galaxies together, we calculate the location of the projected center
of mass of the galaxy `group' and find that it is located at
($+12.3''$, $+21.6''$) from the QSO sightline.  The corresponding
projected distance of the `group' is therefore $d_{\rm group}=118$
pkpc and the \mstar-weighted line-of-sight velocity offset is
$\Delta\,v_{\rm group}=181$ \kms\ between the galaxy `group' and the
LLS.  We also calculate a line-of-sight velocity dispersion of
$\sigma_{\rm group}=230$ \kms\ between these 12 galaxies, implying a
dynamical mass of $M_{\rm dyn}\approx 2\times 10^{13}\,\msun$,
consistent with the LRG halo mass inferred from the mean stellar mass
to halo mass relation (e.g., Behroozi \etal\ 2013).  It is clear that
the LRG (G5) dominates the mass of the `group' and that the
line-of-sight velocity offset between the LLS and the LRG is within
the bound of the halo velocity dispersion.


While our survey has uncovered three previously unknown galaxies at
closer projected distances, $d\apl 100$ pkpc, from the LLS than the
massive LRG, all three galaxies combined contribute to no more than
0.2\% of \mstar\ seen in the LRG at $z=126$ pkpc.  In addition, while
the three group members at $d\apl 100$ pkpc are faint with intrinsic
luminosity $<0.02\,L_*$, the inferred ISM gas phase metallicity of
$> 0.06$ solar is still more than 10 times higher than the 95\%
upper limit derived for the LLS.  In principle, effective mixing
between chemically-enriched ISM ejecta and chemically-pristine IGM can
result in a significant dilution in the metallicity of the LLS (e.g.,
Schaye \etal\ 2007).  To evaluate the effect of chemical dilution, we
first recall the definition of gas metallicity as the mass fraction of
heavy elements, $Z_{\rm gas}\,\equiv\,m({\rm Z})/m_{\rm gas}$.  The
mean gas metallicity of a LLS originating in a cloud of well-mixed ISM
and IGM is therefore
\begin{equation}
\langle\,Z\,\rangle_{\rm LLS}=\frac{Z_{\rm ISM}*m_{\rm ISM}+Z_{\rm IGM}*m_{\rm IGM}}{m_{\rm ISM}+m_{\rm IGM}},
\end{equation}
where $m_{\rm ISM}$ and $m_{\rm IGM}$ represent the total mass of
interstellar and intergalactic gas in the LLS, and the ISM metallicity
is expected to be much greater than what is seen in the IGM, $Z_{\rm
  ISM}\,\gg\,Z_{\rm IGM}$.

Following Equation (1), it is immediately clear that if ISM ejecta
dominates in mass over accreted IGM in the LLS, $m_{\rm
  ISM}\,\gg\,m_{\rm IGM}$, then $\langle\,Z\,\rangle_{\rm
  LLS}\approx\,Z_{\rm ISM}$.  If both ISM ejecta and IGM contribute
comparably to the LLS, $m_{\rm ISM}\,\approx\,m_{\rm IGM}$, then
$\langle\,Z\,\rangle_{\rm LLS}\approx\,1/2\,Z_{\rm ISM}$.  However, if
metal-poor IGM dominates in the LLS, $m_{\rm IGM}\,\gg\,m_{\rm ISM}$,
then $\langle\,Z\,\rangle_{\rm LLS}\,\ll\,Z_{\rm ISM}$ which is what
we see for the LLS at $z=0.33$.

Consequently, the observed disparity in the chemical enrichment level
between the LLS and the group members has two important implications.
First, it is unlikely that the LLS originates in ejecta from these
faint galaxies either by stellar feedback or by ram-pressure stripping
(cf.\ Whiting \etal\ 2006; Kacprzak \etal\ 2010; Gauthier 2013; Bielby
\etal\ 2017).  Observations of nearby low-mass galaxies have
established a mass-metallicity relation of $12+\log\,({\rm O}/{\rm
  H})=(5.61\pm 0.24)+(0.29\pm 0.03)\,\log\,\mstar/\msun$ with a
scatter of $\sigma_{\rm (O/H)}=0.15$ dex over a mass range of
$\log\,\mstar/\msun=7-9.5$ (Berg \etal\ 2012), similar to the mass
range of the low-mass galaxies found near the LLS.  At
$\log\,\mstar/\msun\approx 7$, the ISM metallicity is $\approx 10$\%
solar following this mean relation.  Recently, Hsyu \etal\ (2018)
extended this effort to lower mass with a minimum mass of
$\log\,\mstar/\msun=5.2$.  The mass-metallicity relation of these
ultra-low mass dwarfs is in good agreement with the best-fit relation
of Berg \etal, and remains at $>2.5$\% solar for the lowest-mass
galaxy of $\log\,\mstar/\msun=5.2$ in their sample.  With a lack of
metallicity gradient observed in low-mass galaxies (e.g.,
P{\'e}rez-Montero \etal\ 2016; Belfiore \etal\ 2017), the ISM
metallicity in the outskirts of these dwarf galaxies is likely at a
similar level, still more than a factor of four higher than the LLS.
Condensed cool clumps from the hot halo are also an unlikely
explanation (cf.\ Gauthier \& Chen 2011; Huang \etal\ 2016) under the
expectation that gaseous halos around low-redshift elliptical galaxies
are chemically enriched to solar or near solar metallicities (e.g.,
Mathews \& Brighenti 2003; Mernier \etal\ 2017).  Therefore, the LLS
is likely to originate in infalling gas from the IGM in the LRG halo.
Second, the detection of an extremely low-metallicity LLS in a massive
quiescent halo at $z=0.33$, among the most metal-poor LLS found at $z
< 1$ (e.g., Ribaudo et al.\ 2011), also indicates that parts of the
IGM near old and massive galaxies still remain chemically pristine
through the not too distant past.

While accretion is necessary to sustain galaxy growth, direct
observational evidence of gas accretion remains scarce (e.g., Heitsch
\& Putman 2009; Turner \etal\ 2017; see also Putman 2017 for a recent
review).  Numerical simulations have suggested that gas accretion
proceeds along the disk plane with a limited sky covering fraction of
$\apl 10$\% (e.g., Faucher-Gigu{\`e}re\& Kere{\v s} 2011), making
frequent detections challenging.  Although the study presented here
focuses on a single LLS, the implication is broad.  Recall that the
LLS is one of the seven optically-thick absorbers uncovered from the
COS-LRG program, which was designed to for a systematic survey of the
CGM at $d\apl 160$ pkpc of 16 random LRGs (Paper I).  Three of these
optically-thick absorbers are metal-poor with gas metallicity $<3$\%
solar (Paper II).  Attributing these chemically-pristine LLS to cold
streams accreted from the IGM would imply a mean covering fraction of
$\langle\kappa\rangle\approx 0.18$ for these cold streams, which is
significantly higher than the expectation of few percents for massive
halos at low redshifts from cosmological simulations (cf.\ Hafen
\etal\ 2017).  A larger sample of LLS with deep galaxy survey data
available will be helpful to better quantify the incidence of IGM
accretion galactic halos.

\section*{Acknowledgements}

HWC and FSZ acknowledge partial support from HST-GO-14145.01A and NSF
AST-1715692 grants.  SDJ is supported by a NASA Hubble Fellowship
(HST-HF2-51375.001-A).  SC gratefully acknowledges support from Swiss
National Science Foundation grant PP00P2\_163824.  This work is based
on observations made with ESO Telescopes at the Paranal Observatory
under programme ID 095.A-0200(A), data gathered with the 6.5m Magellan
Telescopes located at Las Campanas Observatory, and imaging data
gathered under the GO-14660 program using the NASA/ESA Hubble Space
Telescope operated by the Space Telescope Science Institute and the
Association of Universities for Research in Astronomy, Inc., under
NASA contract NAS 5-26555.  We acknowledge the use of DECaLS survey
products.

\label{lastpage}

\end{document}